\documentclass[amsmath,12pt]{article}
\usepackage[dvips]{epsfig}
\usepackage{amsmath}

\begin{document}

\title{\bf Electroweak Effects in the Double Dalitz Decay $B_s
  \to l^+ l^- l'^+ l'^-$}
\author{Yusuf Din\c{c}er\footnote{e-mail:dincer@physik.rwth-aachen.de}
  \, and L.M. Sehgal\footnote{e-mail:sehgal@physik.rwth-aachen.de}\\
        Institute of Theoretical Physics, RWTH Aachen \\
        D-52056 Aachen, Germany}
\date{}
\maketitle

\begin{abstract}
We investigate the double Dalitz decays $B_s \to l^+ l^- l'^+ l'^-$ on
the basis of the effective Hamiltonian for the transition $b \bar{s}
\to l^+ l^-$, and universal form factors suggested by QCD.
The correlated mass spectrum of the two lepton pairs  in
the decay $B_s \to e^+ e^- \mu^+ \mu^-$ is derived in an efficient way,
using a QED result for meson decays mediated by two virtual photons:
$B_s \to \gamma^* \gamma^* \to e^+ e^- \mu^+ \mu^-$. A comment is made
on the correlation between the planes of the two lepton  pairs. The
conversion ratios $\rho_{lll'l'}= \frac{ \Gamma(B_s \to l^+ l^- l'^+
  l'^-) }{ \Gamma(B_s \to \gamma \gamma) }$ are estimated to be
$\rho_{eeee}=3 \times 10^{-4}, \rho_{ee\mu\mu}=9 \times 10^{-5} \text{
  and } \rho_{\mu\mu\mu\mu}=3 \times 10^{-5}$, and are enhanced
relative to pure QED by $10-30 \%$.
\end{abstract}

\newpage

\section{Introduction}

In a recent paper \cite{dincersehgal} we investigated the decay $B_s \to l^+ l^-
\gamma \, (l=e,\mu)$, using the effective Hamiltonian for the transition
$b \bar{s} \to l^+ l^-$, and obtained a prediction for the conversion
ratio
\begin{equation}
\rho_{ll} = \frac{ \Gamma (B_s \to l^+ l^- \gamma) }{ \Gamma ( B_s \to
  \gamma \gamma) }
\end{equation}
in terms of the Wilson coefficients $C_7,C_9$ and $C_{10}$. An essential
ingredient of the calculation was the use of a universal form factor
characterising the matrix elements 
$\left< \gamma | \bar{s} i \sigma_{\mu\nu} ( 1+\gamma_5) b | \bar{B}_s
\right>$ and $\left< \gamma | \bar{s} \gamma_\mu ( 1 \pm \gamma_5) b |
  \bar{B}_s \right>$, as suggested by recent work \cite{Korchemsky} on QCD in the
heavy quark limit $(m_b \gg \Lambda_{QCD})$.
It was found that the ratio $\rho_{ll}$ was significantly higher than
one would expect from a QED calculation of Dalitz pair production $B_s
\to \gamma^* \gamma \to l^+ l^- \gamma$, the difference reflecting the
presence of the short-distance coefficients $C_9,C_{10}$, as well as
the universal $1/E_\gamma$ behaviour of the QCD-motivated form
factor. The purpose of the present paper is to apply the same
considerations to the ``double Dalitz decay'' $B_s \to l^+ l^- l'^+
l'^-$, to determine whether there is similar enhancement of the double
conversion ratio
\begin{equation}
\rho_{lll'l'} = \frac{ \Gamma (B_s \to l^+ l^- l'^+ l'^-) }{ \Gamma
  (B_s \to \gamma \gamma) } \,,
\end{equation}
compared to what one would obtain from the QED process $B_s \to
\gamma^* \gamma^* \to l^+ l^- l'^+ l'^-$. We examine also the
correlation in the invariant mass of the two lepton pairs, and the
nature of the angular correlation between the $l^+l^-$ and $l'^+l'^-$
planes, which is a crucial test of the $B_s \to \gamma \gamma$ vertex.

\section{ \bf Matrix Element and Invariant Mass Spectrum}

We begin with the effective Hamiltonian for $b \bar{s} \to l^+ l^-$
\cite{Buchalla}
\begin{eqnarray}\label{effham}
{\cal H}_{\rm eff} &=& \frac{ \alpha G_F}{\sqrt{2}\pi} V_{tb} V^*_{ts} 
\biggl\{
C^{\rm eff}_9        (\bar{s} \gamma_\mu P_L b)         \bar{l}
\gamma_\mu l          \nonumber \\
&+&
C_{10}              (\bar{s} \gamma_\mu P_L b)           \bar{l} \gamma_\mu \gamma_5 l \\
&-&
2 \frac{C_7}{q^2} \bar{s} i \sigma_{\mu\nu} q^\nu (m_b P_R + m_s P_L) b \bar{l} \gamma_\mu l  
\biggr\}  \nonumber
\end{eqnarray}
where $P_{L,R}=(1 \mp \gamma_5)/2$ and $q$ is the sum of the $l^+$ and
$l^-$ momenta. Ignoring small $q^2-$dependent corrections in $C^{\rm
  eff}_9$, the values of the Wilson coefficients are 
\begin{equation} 
C_7 = -0.315, C_9 = 4.334 , C_{10} = -4.624 \,.
\end{equation}
Then, as shown in \cite{dincer}, the matrix element for $\bar{B}_s \to l^+
l^- \gamma$ has the form
\begin{eqnarray}
{\cal M}(\bar{B}_s \to l^+ l^- \gamma) &=& \frac{\alpha
  G_F}{\sqrt{2}\pi} e V_{tb} V^*_{ts} \frac{1}{M_{B_s}} \cdot
\nonumber \\
& & \cdot \left[ \epsilon_{\mu\nu\rho\sigma} \epsilon^{*\nu} q^\rho
  k^\sigma ( A_1 \bar{l} \gamma_\mu l + A_2 \bar{l} \gamma_\mu
  \gamma_5 l ) \right. \\
& & + \left. i ( \epsilon^* (k\cdot q)-(\epsilon^* \cdot q)k_\mu) ( B_1
  \bar{l} \gamma_\mu l+B_2 \bar{l} \gamma_\mu\gamma_5l) \right] \nonumber
\end{eqnarray}
where
\begin{equation}\label{coeff}
\begin{split}
A_1 &= C_9 f_V + 2 C_7 \frac{M^2_{B_s}}{q^2} f_T \,, \\
A_2 &= C_{10} f_V  \,, \\
B_1 &= C_9 f_A + 2 C_7 \frac{M^2_{B_s}}{q^2} f'_T \,, \\
B_2 &= C_{10} f_A \,.
\end{split}
\end{equation}
The form factors $f_V,f_A,f_T,f'_T$, defined in Ref. \cite{dincersehgal}, will be taken
to have the universal form 
\begin{equation}\label{formfactor0}
f_V=f_A=f_T=f'_T = \frac{1}{3} \frac{f_{B_s}}{\Lambda_s}
\frac{1}{x_\gamma} + {\cal O} ( \frac{\Lambda^2_{QCD}}{E^2_\gamma} ) \,,
\end{equation}
predicted in the heavy quark approximation $(m_b \gg
\Lambda_{QCD},m_b \gg m_s)$ in QCD \cite{Korchemsky}. Here,
$\bar{\Lambda}_s=m_{B_s}-m_b \approx 0.5 {\rm GeV}, x_\gamma = 2
E_\gamma/M_{B_s}=1-q^2/M^2_{B_s}$, and $f_{B_s} \approx 200 {\rm MeV}$
is the $B_s$ decay constant. The essential feature for our purpose
will be the universal $1/x_\gamma$ behaviour, the absolute
normalization dropping out in the calculation of the conversion ratio.
(Corrections to universality are discussed in Ref. \cite{Korchemsky1}).

To obtain the matrix element for $B_s \to l^+ l^- l'^+ l'^-$ we treat
the second lepton pair $l'^+l'^-$ as a Dalitz pair associated with
internal conversion of the photon in $B_s \to l^+ l^- \gamma$. From
this point on, we will specialise to the final state
$e^+e^-\mu^+ \mu^-$, consisting of two different lepton pairs. This
avoids the complications due to the exchange diagram that occurs in
dealing with two identical pairs. The matrix element then has the
structure
\begin{equation}\label{matrix}
\begin{split}
{\cal M}(\bar{B}_s \to e^+e^- \mu^+\mu^-) 
&\sim
 \frac{e}{k^2} ( a_+ (q^2)
L_+^\mu (q_1,q_2)+ a_- (q^2) L_-^\mu(q_1,q_2) ) L^\nu_{em}(k_1,k_2) \\
& 
 \cdot [ \epsilon_{\mu\nu\rho\sigma} q^\rho k^\sigma + i(g_{\mu\nu}
k\cdot q- k_\mu q_\nu) ]
\end{split}
\end{equation}
where $k$ and $q$ are the four-momenta of the two lepton pairs, $k^2$
and $q^2$ being the corresponding invariant masses. The currents
$L_\pm$ and $L_{em}$ are given by
\begin{equation}
\begin{split}
L_\pm^\mu (q_1,q_2)  &= \bar{u} (q_1) \gamma^\mu (1 \pm \gamma_5) v
(q_2) \,, \\
L_{\rm em}^\mu (k_1,k_2) &= \bar{u} (k_1) \gamma^\mu v (k_2)  \,.
\end{split}
\end{equation}
where $k_1+k_2=k,q_1+q_2=q$. The coefficients $a_\pm(q^2)$ are related to those in Eq. (\ref{coeff}) by
\begin{equation}
a_\pm (q^2) = A_1(q^2) \pm A_2 (q^2) \,,
\end{equation}
where we have used the fact that for universal form factors, $B_{1,2}
= A_{1,2}.$

At this stage, it is expedient to compare the matrix element (\ref{matrix}) with
the matrix element for double Dalitz pair production in QED. We will
make use of the recent analysis of Barker et al.\cite{Barker}, who have
studied the reaction ${\rm Meson} \to \gamma^* \gamma^* \to l^+ l^-
\l'^+ l'^-$, using a vertex for ${\rm Meson} \to \gamma \gamma$ that
is a general superposition of scalar and pseudoscalar forms, the matrix
element being
\begin{equation}\label{matrixbarker}
\begin{split}
{\cal M}_{\rm Barker} 
&= 
const.\cdot  \frac{e}{k^2} \frac{e}{q^2}
L^\mu_{em} (q_1,q_2) L^\nu_{em}(k_1,k_2) \\
& \cdot  \left[  \xi_P \, \epsilon_{\mu\nu\rho\sigma} q^\rho k^\sigma + \xi_S (
g_{\mu\nu} k \cdot q - k_\mu q_\nu ) \right]  \,.
\end{split}
\end{equation}
The coefficients $\xi_P$ and $\xi_S$ are normalized so that $| \xi_P
|^2 + | \xi_S |^2 = 1$. (In Ref. \cite{Barker} they are denoted by $\xi_P =
\cos \zeta, \xi_S = \sin \zeta e^{i\delta}$.)

From this matrix element, Barker et al. have derived the correlated
invariant mass spectra for the decay into $e^+ e^- \mu^+ \mu^-$
(ignoring form factors at the $M \gamma^* \gamma^*$ vertex)
\begin{equation}\label{resultqed}
\begin{split}
\left[ \frac{1}{\Gamma_{\gamma\gamma}} 
\left( \frac{d^2 \Gamma}{dx_{12} d x_{34}} \right) \right]_{Barker} 
&= 
\left( \frac{2 \alpha^2}{9 \pi^2} \right) \frac{\lambda_{12} \lambda_{34} \lambda}{w^2}
(3-\lambda^2_{12})(3-\lambda^2_{34})  \\
& \cdot 
 \left[ |\xi_P|^2 \lambda^2 + |\xi_S|^2 (\lambda^2 + \frac{3 w^2}{2})
 \right] \,.
\end{split}
\end{equation}
The variables entering
the above formula are defined as follows:
\begin{equation}
\begin{split}
x_{12} &= (q_1+q_2)^2/M^2 = q^2/M^2  \,, \\
x_{34} &= (k_1+k_2)^2/M^2 = k^2/M^2  \,, \\
x_1 &= x_2 = \frac{m^2_1}{M^2} \frac{1}{x_{12}} \,,  \\
x_3 &= x_4 = \frac{m^2_3}{M^2} \frac{1}{x_{34}} \,, \\
z &= 1-x_{12}-x_{34} \,, \\
\lambda_{12} &= \sqrt{(1-x_1-x_2)^2-4 x_1 x_2} \,, \\
\lambda_{34} &= \sqrt{(1-x_3-x_4)^2-4 x_3 x_4} \,, \\
w^2 &= 4 x_{12} x_{34} \,, \\ 
\lambda &= \sqrt{z^2-w^2} \,.
\end{split}
\end{equation}
Here $m_1$ and $m_3$ denote the masses of the electron and muon, and
$M$ the mass of  the decaying meson. The phase space in the variables
$x_{12}$ and  $x_{34}$ is defined by $x^0_{34} < x_{34} <
(1-\sqrt{x_{12}})^2, x^0_{12} < x_{12} < (1-\sqrt{x_{34}})^2$, where
$x^0_{12} = 4 m^2_1/M^2, x^0_{34} = 4 m^2_3/M^2$.

We can now adapt the QED result  (\ref{resultqed}) to the process $B_s \to e^+ e^-
\mu^+ \mu^-$, by comparing the matrix element (\ref{matrixbarker}) with that in Eq.
(\ref{matrix}).  The essential observation is that in the approximation of
neglecting lepton masses, the vector and axial vector parts of the
chiral currents $L^\mu_\pm$ contribute equally and independently to the
invariant mass spectrum. In addition, the matrix element for $B_s$
decay corresponds to the QED matrix element considered by Barker et
al., if we put $ \xi_P  = 1/\sqrt{2} , \xi_S =i/\sqrt{2}$. This allows us to obtain the
invariant mass spectrum for the double Dalitz decay  $B_s \to e^+
e^- \mu^+ \mu^-$ in electroweak theory:
\begin{equation}\label{dgammaew}
\begin{split}
\left[ \frac{1}{\Gamma_{\gamma\gamma}} \left( \frac{d \Gamma}{d x_{12}
      d x_{34}} \right) \right]_{EW}
&=
\left\{ \left[ (\eta_9+\frac{1}{x_{12}})^2 + \eta^2_{10} \right] + \left[ (\eta_9 +
    \frac{1}{x_{34}})^2 + \eta^2_{10} \right] 
       \right\} \\
& \cdot 
 \frac{x^2_{12} x^2_{34}}{x^2_{12}+x^2_{34}} \left| F(x_{12},x_{34}) \right|^2 \cdot 
\left[ \frac{1}{\Gamma_{\gamma\gamma}} \left( \frac{d \Gamma}{d x_{12}
      d x_{34}} \right) \right]_{QED} \,,
\end{split}
\end{equation}
where 
\begin{equation}
\left[ \frac{1}{\Gamma_{\gamma\gamma}} \left( \frac{d \Gamma}{d x_{12}
      d x_{34}} \right) \right]_{QED} = \frac{\alpha^2}{9\pi^2}
\frac{\lambda_{12} \lambda_{34} \lambda}{w^2} (3-\lambda_{12}^2)
(3-\lambda_{34}^2)  (2 \lambda^2 + \frac{3}{2} w^2 ) \,.
\end{equation}
Here we have used the abbreviation $\eta_9 = C_9/(2C_7)$ and
$\eta_{10} = C_{10}/(2C_7)$, introduced in Ref. \cite{dincersehgal}.
The electroweak formula (\ref{dgammaew}) reduces to the QED result in
the limit $\eta_9=\eta_{10}=0, F(x_{12},x_{34})=1$.

The form factor $F(x_{12},x_{34})$  is chosen to have the universal form
\begin{equation}\label{formfactor}
F(x_{12},x_{34}) = \frac{1}{(1-x_{12})} \frac{1}{(1-x_{34})} \,.
\end{equation}
(a possible normalization factor drops out in the calculation of the
conversion ratio).
This is a plausible (but not unique) generalization of the universal QCD
form factor $1/(1-x_{12})$ that occurs in the single Dalitz pair
process $B_s \to e^+ e^- \gamma$.

\begin{figure}
  \epsfig{file=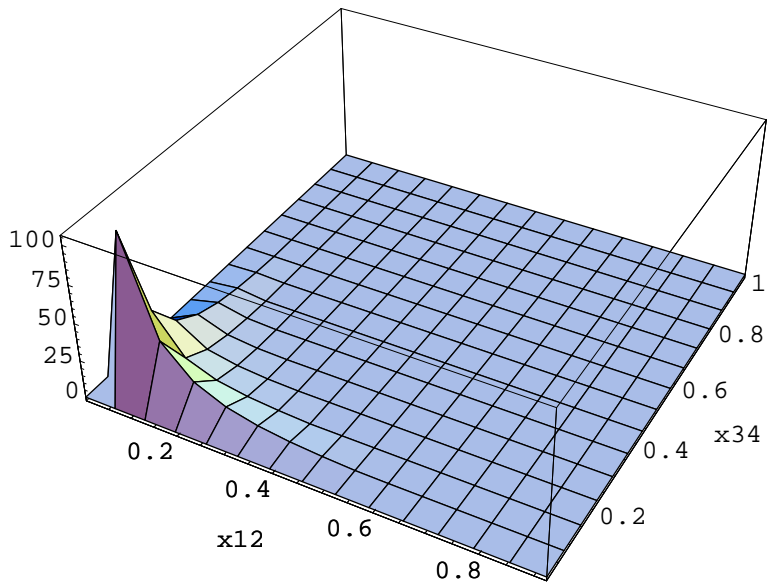}
  \caption{Invariant mass distribution $d\Gamma/d x_{12} d x_{34}$ for
    $B_s \to e^+ e^- \mu^+ \mu^-$ in electroweak theory}
  \label{plot1}
\end{figure}

\begin{figure}
  \epsfig{file=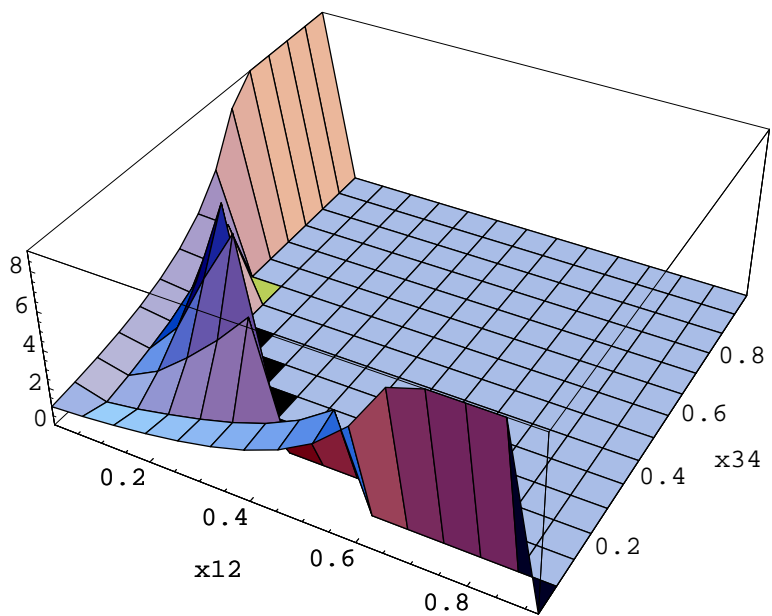}
  \caption{Ratio $(d\Gamma/d x_{12} d x_{34})_{EW}/(d\Gamma/d x_{12} d
    x_{34})_{QED}$ showing influence of electroweak parameter
    $C_9,C_{10}$ and the form factor $F(x_{12},x_{34})$.}
  \label{plot2}
\end{figure}

\begin{figure}
  \epsfig{file=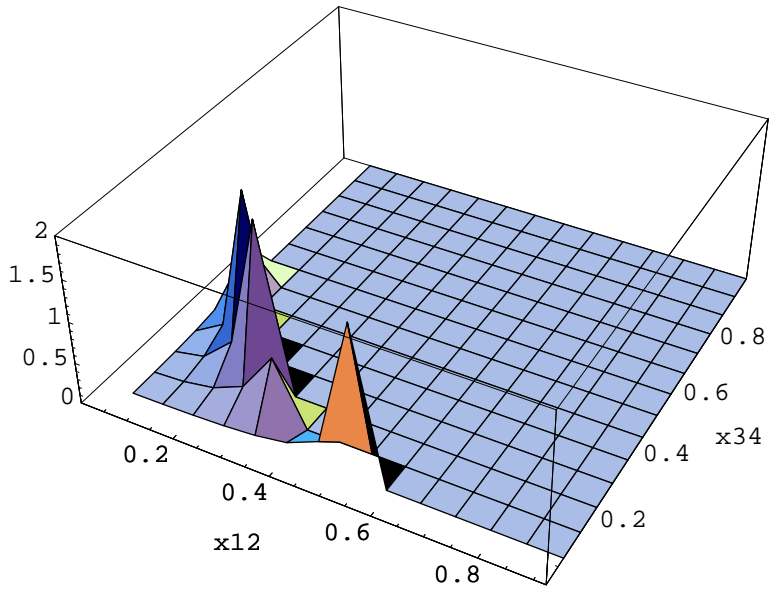}
  \caption{Ratio $(d\Gamma/d x_{12} d x_{34})_{EW}/(d\Gamma/d x_{12} d
    x_{34})_{QED}$ in the limit of a constant form factor $(
    F(x_{12},x_{34})=1)$, illustrating specific effect of electroweak
    parameters $\eta_9=C_9/(2C_7)$ and $\eta_{10}=C_{10}/(2C_7)$.}
  \label{plot3}
\end{figure}

In Fig. \ref{plot1} we plot the correlated invariant mass spectrum for $B_s \to
e^+ e^- \mu^+ \mu^-$ in electroweak theory. 
The ratio of the electroweak and QED spectra is shown in Fig \ref{plot2}, and
indicates the effects associated with the coefficients $\eta_9$ and
$\eta_{10}$, and the form factor $F(x_{12},x_{34})$. One notes a
slight depression in the region $x_{12}=- \frac{2 C_7}{C_9}$ or $x_{34}
= - \frac{2 C_7}{C_9}$, connected with the vanishing of the term
$(C_9+\frac{2 C_7}{x_{12}})^2$ or $(C_9+\frac{2
  C_7}{x_{34}})^2$. There is also a general enhancement for increasing
values of $x_{12},x_{34}$, because of the form factor
(\ref{formfactor}). If the form factor $F(x_{12},x_{34})$ is set equal
to one, the ratio of the electroweak and QED spectra has the structure
plotted in Fig. \ref{plot3}, illustrating the effects which depend
specifically on the
electroweak parameters $\eta_9,\eta_{10}$.

The absolute value of the conversion ratio $\rho_{ee\mu\mu}$ is
obtained by integrating \\ $( \frac{1}{\Gamma_{\gamma\gamma}} d \Gamma/d x_{12} d x_{34})_{EW}$ over the
range of $x_{12}$ and $x_{34}$.
In the QED case, this ratio is conveniently expressed in
terms of the integrals $I_{1\ldots 6}$ introduced in Ref. \cite{Barker}:

\begin{equation}
\begin{split}
I_1 &= \frac{2}{3} \int \int d x_{12} d x_{34}  \frac{
  \lambda^3_{12} \lambda^3_{34} \lambda^3}{w^2} \,, \\
I_2 &= \frac{2}{3} \int \int d x_{12} d x_{34}  \frac{
  \lambda^3_{12} \lambda^3_{34} \lambda z^2}{w^2} \,, \\
I_3 &= \frac{4}{3} \int \int d x_{12} d x_{34}  \frac{
  \lambda^3_{12} \lambda^3_{34} \lambda^2 z}{w^2} \,, \\
I_4 &= \int \int d x_{12} d x_{34}  \frac{ \lambda_{12}
  \lambda_{34} \lambda^3}{w^2} (3-\lambda^2_{12} - \lambda^2_{34}) \,,\\
I_5 &= \int \int d x_{12} d x_{34}  \frac{\lambda_{12}
  \lambda_{34} \lambda z^2}{w^2} (3-\lambda^2_{12} -\lambda^2_{34})\,, \\
I_6 &= \frac{1}{6} \int \int d x_{12} d x_{34}  \lambda_{12}
\lambda_{34} \lambda (3-\lambda^2_{12})(3-\lambda^2_{34}) \,.
\end{split}
\end{equation}
These integrals are listed in Table \ref{integrals} (where, for
completeness, we have also given the values for the final states
$e\bar{e}e\bar{e}$ and $\mu \bar{\mu} \mu \bar{\mu}$).
These integrals allow us to calculate the QED double
conversion ratio
\begin{equation}\label{ratioqed}
\begin{split}
(\rho_{ee\mu\mu})_{QED} &= \frac{\alpha^2}{6 \pi^2} ( I_1 + I_2 + 2
(I_4 + I_5 + I_6) ) \\
& = 7.6 \times 10^{-5} \,.
\end{split}
\end{equation}
The corresponding result for electroweak theory, based on the
differential decay rate (\ref{dgammaew}), can be expressed in terms of the
integrals
\begin{equation}\label{tildeintegral}
\begin{split}
\tilde{I}_1 &= \frac{2}{3} \int \int d x_{12} d x_{34}  \frac{
  \lambda^3_{12} \lambda^3_{34} \lambda^3}{w^2} G(x_{12},x_{34})\,, \\
\tilde{I}_2 &= \frac{2}{3} \int \int d x_{12} d x_{34}  \frac{
  \lambda^3_{12} \lambda^3_{34} \lambda z^2}{w^2} G(x_{12},x_{34})\,, \\
\tilde{I}_3 &= \frac{4}{3} \int \int d x_{12} d x_{34}  \frac{
  \lambda^3_{12} \lambda^3_{34} \lambda^2 z}{w^2} G(x_{12},x_{34}) \,, \\
\tilde{I}_4 &= \int \int d x_{12} d x_{34}  \frac{ \lambda_{12}
  \lambda_{34} \lambda^3}{w^2} (3-\lambda^2_{12} - \lambda^2_{34})
G(x_{12},x_{34}) \,,\\
\tilde{I}_5 &= \int \int d x_{12} d x_{34}  \frac{\lambda_{12}
  \lambda_{34} \lambda z^2}{w^2} (3-\lambda^2_{12} -\lambda^2_{34})
G(x_{12},x_{34}) \,, \\
\tilde{I}_6 &= \frac{1}{6} \int \int d x_{12} d x_{34}  \lambda_{12}
\lambda_{34} \lambda (3-\lambda^2_{12})(3-\lambda^2_{34})
G(x_{12},x_{34}) \,.
\end{split}
\end{equation}
The factor $G(x_{12},x_{34})$ in the integrand of
Eq.(\ref{tildeintegral}) contains the effects of the electroweak
coefficients $\eta_9,\eta_{10}$ and the universal form factor
$F(x_{12},x_{34})$:
\begin{equation}
\begin{split}
G(x_{12},x_{34}) &= \left\{ \left[ \left( \eta_9+\frac{1}{x_{12}}
    \right)^2 + \eta_{10}^2 \right] + \left[ 
\left( \eta_9+\frac{1}{x_{34}} \right)^2 + \eta_{10}^2 \right] 
\right\} \cdot \\
& \cdot \frac{x^2_{12} x^2_{34}}{x^2_{12}+x^2_{34}} \cdot
|F(x_{12},x_{34})|^2 \,.
\end{split}
\end{equation}
The integrals $\tilde{I}_1, \ldots, \tilde{I}_6$ are given in Table
\ref{integrals2}. The electroweak conversion ratio, analogous to the
QED result (\ref{ratioqed}), is given by
\begin{equation}\label{ratioew}
\begin{split}
( \rho_{ee\mu\mu})_{EW} &= \frac{\alpha^2}{6\pi^2} \left( \tilde{I}_1 +
  \tilde{I}_2 + 2 (  \tilde{I}_4 + \tilde{I}_5 + \tilde{I}_6) \right)
\\
& = 9.1 \times 10^{-5} \,.
\end{split}
\end{equation}
In comparison to the QED result (\ref{ratioqed}), the double
conversion ratio for $B \to e \bar{e} \mu \bar{\mu}$ in electroweak
theory is enhanced by $\sim 20 \%$.

A calculation of the spectra for the channels $e\bar{e} e \bar{e}$ and
$\mu\bar{\mu} \mu \bar{\mu}$ is complicated by interference between
the exchange and direct amplitudes.
The conversion ratio for these channels takes the form
\begin{equation}
\rho = \rho_1 + \rho_2 +\rho_{12} \,,
\end{equation}
where $\rho_1$ and $\rho_2$ denote the ``direct'' and ``exchange''
contribution, and $\rho_{12}$ an interference term. Numerical
calculations of the decays $\pi^0 \to e^+ e^- e^+ e^-$ and $K_L \to
e^+ e^- e^+ e^-$ suggest that $\rho_{12}$ is small and $\rho_1 \approx
\rho_2$. Thus a rough estimate of the double conversion ratio can be
obtained using the formula (\ref{ratioew}), with an extra factor
$(\frac{1}{4})\cdot2$ where $(\frac{1}{4})$ is the statistical factor
for two identical fermion pairs, and $2$ comes from adding direct and
exchange contributions. This yields, using the numbers in \\ Table \ref{integrals2}
\begin{equation}
\begin{split}
( \rho_{e\bar{e}e\bar{e}} )_{EW} & \approx 2.9 \times 10^{-4} \,, \\
( \rho_{\mu \bar{\mu} \mu \bar{\mu}} )_{EW} & \approx 2.8 \times
10^{-5} \,.
\end{split}
\end{equation}
For comparison, the QED results, using Table \ref{integrals},
are
\begin{equation}
\begin{split}
( \rho_{e\bar{e}e\bar{e}} )_{QED} & \approx 2.7 \times 10^{-4} \,, \\
( \rho_{\mu \bar{\mu} \mu \bar{\mu}} )_{QED} & \approx 2.2 \times 10^{-5} \,.
\end{split}
\end{equation}
Thus the enhancement in the case of $e\bar{e} e \bar{e}$ is $\sim
10\%$ and that in $\mu\bar{\mu} \mu\bar{\mu}$ about $30\%$.
Combining (21) and (23),  the ratio of the
channels $e\bar{e}e\bar{e},e\bar{e}\mu\bar{\mu}$ and
$\mu\bar{\mu}\mu\bar{\mu}$ is approximately
\begin{equation}
\begin{split}
e\bar{e}e\bar{e} &: e\bar{e}\mu\bar{\mu} : \mu\bar{\mu}\mu\bar{\mu} \\
= \quad 3 &: 1 : 0.3 
\end{split}
\end{equation}

To obtain the absolute branching ratios, we note that the 
decay rate of $\bar{B}_s \to \gamma \gamma$, derived from the
effective Hamiltonian (\ref{effham}), involves the Wilson
coefficient $C_7$ and the universal form factor $f_T(x_\gamma=1)$ (see
Eq. (\ref{formfactor0})). Using nominal values for $f_{B_s}$ and $\bar{\Lambda}_s$,
and evaluating $C_7$ at the renormalization scale $\mu=m_b$,
Ref. \cite{Bosch} finds ${\rm Br}(B_s \to \gamma \gamma)=1.23 \times
10^{-6}$. Using this as a reference value, we obtain:
\begin{equation}
\begin{split}
{\rm Br} (\bar{B}_s \to e \bar{e} e \bar{e}) &= 3.6 \times 10^{-10}
\,, \\
{\rm Br} (\bar{B}_s \to e \bar{e} \mu \bar{\mu}) & =  1.1 \times
10^{-10} \,, \\
{\rm Br} (\bar{B}_s \to \mu \bar{\mu} \mu \bar{\mu}) & = 3.5 \times
10^{-11} \,.
\end{split}
\end{equation}

\section{\bf Correlation of $e^+e^-$ and $\mu^+\mu^-$ planes in
  $\bar{B}_s \to e \bar{e} \mu \bar{\mu}$}

One of the distinctive features of the electroweak $\bar{B}_s \to
\gamma \gamma$ matrix element is that the coefficients $\xi_S$ and $\xi_P$
(normalized to $|\xi_S|^2+|\xi_P|^2=1$) are given by $\xi_S =
\frac{i}{\sqrt{2}}$ and $\xi_P = \frac{1}{\sqrt{2}}$. The equality
$|\xi_S|^2=|\xi_P|^2$ leads to the simplification that the factor $|\xi_P|^2
\lambda^2 + |\xi_S|^2 ( \lambda^2 + \frac{3}{2} w^2)$ appearing in the
spectrum (\ref{resultqed}) could be written as
$\frac{1}{2} [ 2 \lambda^2 + \frac{3}{2} w^2]$ in going over to the
electroweak case (Eq.(15)). A further interesting consequence is 
the distribution of the angle $\phi$ between the $e^+e^-$ and $\mu^+
\mu^-$ planes in $\bar{B}_s \to e \bar{e} \mu
\bar{\mu}$. Generalising the QED result given in Ref. \cite{Barker} to the
electroweak case, the correlation in $\phi$ is given by
\begin{equation}
\left( \frac{1}{\Gamma_{\gamma\gamma}} \frac{d \Gamma}{d \phi} 
\right)^{e \bar{e} \mu \bar{\mu}}_{EW} = \frac{\alpha^2}{6 \pi^3} 
\left[ \tilde{I}_1 \sin^2 \phi + \tilde{I}_2 \cos^2 \phi + ( 
  \tilde{I}_4 + \tilde{I}_5 + \tilde{I}_6) \right] \,.
\end{equation} 
The fact that $\tilde{I}_2$ is so close to $\tilde{I}_1$ means that the
spectrum $d \Gamma/d \phi$ is essentially independent of
$\phi$. Furthermore, the fact that $arg(\xi_S/\xi_P)=\pi/2$ refelects
itself in the absence of a term proportional to $\sin \phi \cos \phi$,
the presence of which would lead  to an asymmetry between events with
$\sin \phi \cos \phi > 0$ and $<0$.

It may be remarked that there are corrections to the $B_s \to \gamma
\gamma$ matrix element (associated, for example, with the elementary
process $b \bar{s} \to c \bar{c} \to \gamma \gamma$) which cause the
superposition of scalar and pseudoscalar terms to deviate slightly
from the ratio $\xi_S/\xi_P=i$ \cite{Bosch,Chang}. From the work of Bosch and Buchalla
\cite{Bosch}, we find
\begin{equation}
\frac{\xi_S}{\xi_P} = i \left[ 1 - \frac{2}{3} \frac{C_1+N C_2}{C_7}
  \frac{\lambda_B}{m_B} g(z_c) \right]^{-1}
\end{equation}
where
\begin{equation}
g(z) \approx -2 + ( -2 \ln^2 z + 2 \pi^2 - 4 \pi i \ln z ) z + {\cal
  O}(z^2) \,,
\end{equation}
and $z_c = m_c^2/m^2_b \sim 0.1, C_1 = 1.1, C_2= -0.24, N=3$.
There is thus a small correction to the equality $|\xi_P|=|\xi_S|$. More
interestingly, the phase $\delta= arg(\xi_P/\xi_S)$ is not exactly
$90^0$,implying that a term of the form $\tilde{I}_3 \sin \phi \cos
\phi \cos \delta$ could appear in $d \Gamma/ d \phi$. These corrections are,
however, too small, to have a measurable impact on the spectrum and
branching ratio of the  decay $B_s \to e \bar{e} \mu \bar{\mu}$
calculated above.

\section{\bf Conclusions}

We have calculated the spectrum and rate of the double Dalitz decay
$\bar{B}_s \to e^+ e^- \mu^+ \mu^-$, using the effective Hamiltonian
for the flavour-changing neutral current reaction $b \bar{s} \to l^+
l^-$, and form-factors motivated by the heavy quark limit of QCD. A
method is given for obtaining the correlated mass spectrum $d \Gamma/d
x_{12} d x_{34}$ from the known results for the QED process
  $\bar{B}_s \to \gamma^* \gamma^* \to e^+ e^- \mu^+ \mu^-$. The
  conversion ratios $\rho_{l\bar{l}l' \bar{l}'} = \Gamma(B_s \to l^+
  l^- l'^+ l'^-) / \Gamma (B_s \to \gamma \gamma)$ show an enhancement
  over the QED result, ranging from $10\%$ for the channel $e^+e^-
 e^+ e^-$ to $30\%$ for the channel $\mu^+ \mu^- \mu^+ \mu^-$.
Our best estimate of the branching ratios, using the QCD estimate
${\rm Br} (B_s \to \gamma \gamma) = 1.23 \times 10^{-6}$ given in
\cite{Bosch}, is ${\rm Br} (B_s \to e \bar{e}e \bar{e})= 3.6 \times 10^{-10}$,
${\rm  Br} (B_s \to e \bar{e} \mu \bar{\mu}) = 1.1 \times 10^{-10}, 
 {\rm Br} (B_s \to  \mu \bar{\mu}\mu
\bar{\mu}) = 3.5 \times 10^{-11}$. These
branching ratios may have a chance of being observed at future hadron machines producing
up to $10^{12}$ $B_s$ mesons.

\section*{\bf Acknowledgments}

We wish to thank A. Chapovsky for a useful discussion.
One of us (Y.D.) acknowledges the award of a Doctoral stipend from the
state of Nordrhein-Westphalen.

\newpage

\begin{table}
\begin{tabular}{||c|c|c|c||}
\hline \hline
 & $B_s \to e e \mu \mu$ & $B_s \to e e e e $ & $B_s \to \mu \mu \mu
 \mu$ \\
\hline \hline
 & & & \\
$I_1$ & 7.754 & 32.501 & 1.772 \\
\hline
&&& \\
$I_2$ & 7.806 & 32.556 & 1.821 \\
\hline
&&&\\
$I_3$ & 15.556 & 65.053 & 3.589 \\
\hline
&&&\\
$I_4$ & 17.558 & 58.416 & 5.115 \\
\hline
&&&\\
$I_5$ & 17.641 & 58.499 & 5.199 \\
\hline
&&&\\
$I_6$ & 0.0548 & 0.0556 & 0.0540 \\
\hline \hline
\end{tabular}
\caption{Numerical values of the integrals $I_1,\ldots,I_6$ for $B_s
  \to l \bar{l} l' \bar{l}'$ in QED}
\label{integrals}
\end{table}

\begin{table}
\begin{tabular}{||c|c|c|c||}
\hline \hline
 & $B_s \to e e \mu \mu$ & $B_s \to e e e e $ & $B_s \to \mu \mu \mu
 \mu$ \\
\hline \hline
&&& \\
$\tilde{I}_1$ & 9.336 & 35.491 & 3.856\\
\hline
&&& \\
$\tilde{I}_2$ & 9.477 & 35.643 & 4.002 \\
\hline
&&& \\
$\tilde{I}_3$ & 18.793 & 71.114 & 7.837 \\
\hline
&&& \\
$\tilde{I}_4$ & 20.411 & 63.457 & 5.784 \\
\hline
&&& \\
$\tilde{I}_5$ & 20.637 & 63.685 & 6.003 \\
\hline
&&& \\
$\tilde{I}_6$ & 0.148 & 0.152 & 0.146 \\
\hline \hline
\end{tabular}
\caption{Numerical values of the integrals
  $\tilde{I}_1,\ldots,\tilde{I}_6$ for $B_s \to l \bar{l} l' \bar{l}'$
  in electroweak theory}
\label{integrals2}
\end{table}

\end{document}